**An Iterative Algorithm to Impute Truck Information over Nationwide Traffic Networks[1]**


Diyi Liu[a*], Ankur Shiledar[b], Hyeonsup Lim[c], Vivek Sujan[c], Adam Siekmann[c], Junchuan Fan[c], Lee D. Han[a]

[a]School of Civil and Environmental Engineering, University of Tennessee, Knoxville, Tennessee, USA
[b]School of Civil and Environmental Engineering, Ohio State University, Ohio, USA
[c]Buildings and Transportation Science Division, Oak Ridge National Laboratory, Oak Ridge, Tennessee, USA
[*]Corresponding author: Diyi Liu (dliu27@vols.utk.edu)


---


[1] Notice: This manuscript has been authored by UT-Battelle, LLC, under contract DE-AC05-00OR22725 with the US Department of Energy (DOE). The US government retains and the publisher, by accepting the article for publication, acknowledges that the US government retains a nonexclusive, paid-up, irrevocable, worldwide license to publish or reproduce the published form of this manuscript, or allow others to do so, for US government purposes. DOE will provide public access to these results of federally sponsored research in accordance with the DOE Public Access Plan.





**Abstract**

Understanding the dynamics of truck volumes and activities across the skeleton traffic network is pivotal for effective traffic planning, traffic management, sustainability analysis, and policy making. Yet, relying solely on average annual daily traffic volume for trucks cannot capture the temporal changes over time. Recently, the Traffic Monitoring Analysis System dataset has emerged as a valuable resource to model the system by providing information on an hourly basis for thousands of detectors across the United States. Combining the average annual daily traffic volume from the Highway Performance Monitoring System and the Traffic Monitoring Analysis System dataset, this study proposes an elegant method of imputing information across the traffic network to generate both truck volumes and vehicle class distributions. A series of experiments evaluated the model's performance on various spatial and temporal scales. The method can be helpful as inputs for emission modeling, network resilience analysis, etc.

1. **Introduction**

Understanding the dynamics of freight flows is vital not only for effective transportation planning and management but also for assessing their implications on energy consumption and environmental sustainability. Recent advancements in the field have yielded a suite of tools tailored for the analysis of freight emissions analysis (Shiledar et al., 2024). Additionally, the emergence of electric vehicles has been touted for their potential economic and environmental advantages, particularly in energy conservation (Weiss et al., 2020). Consequently, there exists a pressing demand to systematically evaluate the efficacy of various policies and investments within the trucking and freight industries. However, a considerable portion of existing research relies on synthetic, simulated, or proprietary trucking data, underscoring the necessity to develop a data-centric approach to comprehensively model truck flows and activities over the traffic network.

To analyze traffic patterns in the United States, researchers often use the Highway Performance Monitoring System (HPMS) dataset, a widely recognized source of information. Alongside fundamental roadway characteristics, the US Department of Transportation mandates the reporting of annual average daily traffic (AADT) figures annually. Ideally, a comprehensive assessment would involve daily traffic counts at each location throughout the year to derive accurate average traffic volumes. However, practical constraints often limit this approach, particularly for smaller roads where AADT values are typically approximated through a 3 day observation period within a given year. Usually, a limited number of detectors are rotated across smaller road segments over the course of the year. Subsequently, the observed values undergo correction processes to determine AADT, accounting for various factors such as seasonal variations and weekday fluctuations. AADT values serve as crucial metrics for federal, state, and



metropolitan planning agencies (Krile et al., 2016). Nonetheless, it is still necessary to enhance the precision of daily, weekly, and monthly traffic fluctuation modeling through more refined observational and analytical techniques.

To address the imperative for a more detailed nationwide traffic flow database, the Traffic Monitoring Analysis System (TMAS) dataset has been established to systematically monitor traffic patterns across the United States ("2022 Traffic Monitoring Guide," 2022; Krile, 2016). Predominantly derived from sensors strategically positioned along major highways or arterials, the TMAS dataset comprises three primary subsets: (1) the volume dataset, detailing the count of vehicles traversing a designated point along a roadway in a consistent direction; (2) the class dataset, elucidating the distribution of vehicle classes; and (3) the weight dataset, delineating the distribution of vehicle weights.

**Figure 1** illustrates the geographical distribution of traffic observation stations according to the TMAS dataset compiled in 2021. The dataset comprises approximately 4,558 stations for measured traffic volume and 3,573 stations for vehicle class assessments. Each yellow dot depicted in **Figure 1** (a) represents a site for volume observation, whereas each green dot in **Figure 1** (b) indicates a site for vehicle class observation. The geographic locations of weight stations are not shown because they are out of the scope of this study. For this study, only the class datasets were used for analysis, providing insights into traffic volume across the Federal Highway Administration's 13 vehicle classes on an hourly basis (Pagan-Ortiz, 2014). Note that some states did not report class datasets to TMAS for 2021, but all states uploaded their volume information.



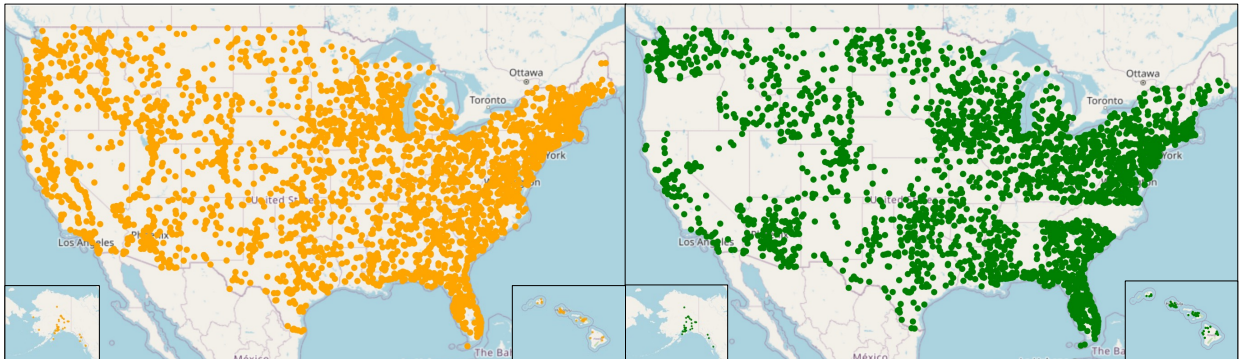

**Figure 1. Locations of TMAS detectors for both (a) volume observation and (b) vehicle class observation.**

Although TMAS data offer extensive coverage of geographical areas, their recordings are confined to predetermined locations or points at the traffic networks. In this investigation, our aim was to estimate traffic volume across the entire traffic network by leveraging TMAS recordings from fixed points. Specifically, the research team conducted imputation on the Freight Analysis Framework (FAF) networks as delineated in previous studies (Hwang et al., 2021, 2016; Department of Transportation et al., 2017). Additionally, we integrated AADT truck volume reported from the HPMS into the FAF network as weighting factors to propagate information over the network.

**Figure 2** illustrates a simplified example to demonstrate the process of imputation on a grid network with directed edges, originating from the lower left corner and extending to the upper right corner. Truck percentages are set at 10% and 100% at the starting and ending edges, respectively. Initially, all nodes within the grid network, except for the starting and ending edges, lack information. Using an iterative algorithm, missing data within the grid network can be gradually imputed by iteratively running the algorithm multiple times. As depicted in the center of **Figure 2**, which represents 100 iterations, the proportion of trucks steadily transitions from



10% to 100%. Subsequently, the results converged after about 200 iterations, as shown on the right of **Figure 2**. Through the implementation of the algorithm, inferring the distribution of vehicle classes becomes feasible.

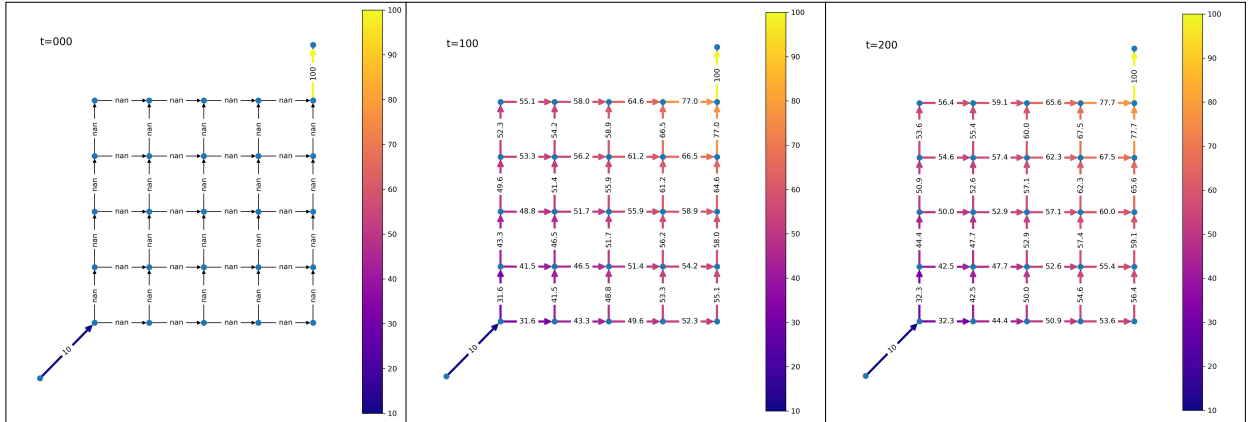

**Figure 2. A simple example of imputing missing traffic volume in a grid network traveling from bottom left to top right.**

The remainder of this work is structured as follows: **Section 2** provides a comprehensive review of prior methodologies and approaches in traffic volume estimation. **Section 3** presents the formulation of the proposed algorithm along with associated data processing steps necessary to facilitate its implementation. **Section 4** presents the background and outcomes of a series of experiments conducted to assess efficacy of the algorithm from diverse analytical viewpoints. Finally, **Section 5** concludes this study by delving into its contributions, potential implications, inherent limitations, and avenues for future research.

2. Literature Review

Traffic detectors are conventionally positioned at various intervals along roadways, necessitating the estimation or interpolation of traffic data across the entire network. Prior studies have predominantly used statistical methodologies to forecast the AADT values of roads, drawing upon factors such as roadway characteristics, socioeconomic indicators, and proximity to highways (Anderson et al., 2006; Caceres et al., 2018; Chang and Cheon, 2019;



Mohamad et al., 1998; Wang and Kockelman, 2009; Zhao and Chung, 2001). These regression techniques, termed aspatial methods, neglect the topology of the traffic network, thereby presenting certain limitations. Moreover, to improve performance, these methodologies need special socioeconomic data collection for each localized analysis zone within a metropolitan area or city. They rely solely on statistical modeling by using input features and do not integrate roadway geometries or network topologies.

In addition to the aspatial methods, various approaches exist for estimating roadway information by leveraging data from neighboring roadways. One prominent technique is geographically weighted regression (GWR), which entails computing a weighted average that considers geographical proximity. GWR has widespread application across diverse domains, including traffic safety analysis and traffic management (Gu et al., 2023; Hezaveh et al., 2019; Xu et al., 2020), and has been adapted for estimating AADT values (Liu et al., 2019; Zhao and Park, 2004). The limitation of using the GWR for estimating AADT is that its performance might not be as effective in urban areas as opposed to rural areas.

Besides the GWR method, Kriging methods are also employed for traffic flow imputation. Originally devised for spatial interpolation of mining resources based on sampled points (Gotway et al., 1996), Kriging has been adapted for transportation applications to interpolate values along traffic corridors and networks (Lowry, 2014). An extended Kriging method is used in transportation studies to impute values in spatiotemporal dimensions across traffic networks (Bae et al., 2018; Yang et al., 2018).



More recently, graph neural networks have gained traction for modeling traffic based on network structures in various applications, including traffic prediction (Bui et al., 2022; Jiang and Luo, 2022; Li and Zhu, 2021), data imputation (Kong et al., 2023; Liang et al., 2022; Shen et al., 2023), event identification (Zhang et al., 2020), and so forth. Despite claims of state-of-the-art performance, graph neural networks exhibit limitations, notably a focus on metropolitan areas rather than statewide or nationwide analyses, in which detector densities may be insufficient to adequately capture the roadway geometry. Moreover, training such models at state or national levels may become very computationally intensive.

Considering the constraints elucidated in previous studies, the reminder of this work will introduce the proposed methodology and analytical framework with the following objectives: (1) to devise an efficient method for imputing traffic data throughout a traffic network, (2) to construct a versatile analytical framework capable of data imputation at varying temporal aggregation levels as desired, and (3) to thoroughly analyze the advantages and limitations of the model across different scenarios and potential usages.

## 3. Methodology

### 3.1 An overview of the data analysis framework

To systematically address the issue, we proposed an analytical framework comprising four primary stages, as shown in **Figure 3**. The first step was the generation of AADT values and their alignment with the FAF traffic network. The second step was the delimitation of the temporal scope of interest and aggregation of TMAS class datasets to acquire freight volume and class distribution insights at TMAS locations. Additionally, the second step included the correlation of these data with the nearest directional link within the FAF traffic network. The



third step included comprehensive integration of all gathered information to impute TMAS data across the entire network for both volume and class distribution. Finally, the fourth step included implementation of cross validation techniques involving the masking of selected TMAS stations as validation sets, followed by a comparison of predicted values against observed ones within these validation sets. Furthermore, the study's outcomes could extend beyond the discussed parameters, offering potential benefits for various ancillary analyses such as emission modeling and activity pattern analysis through the provision of these more informed, data-driven estimations.

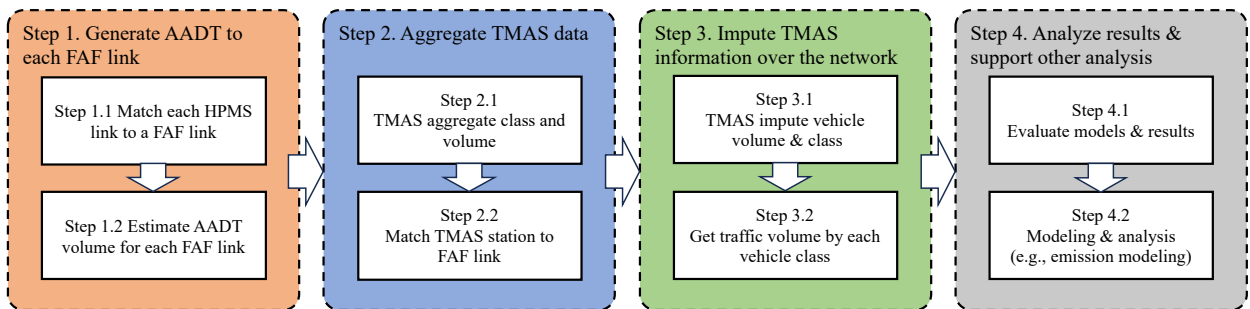

**Figure 3. The proposed analysis framework for estimating vehicle volume and class distributions along traffic networks based on AADT values and TMAS datasets.**

Integration of AADT values and TMAS information into a cohesive network representation is imperative for effective imputation. We employed the FAF network to amalgamate these datasets to form one graph representing all information. Although the HPMS network is another available option, the FAF network offers a more realistic geometric framework characterized by intricate geometries alongside a streamlined set of traffic links.

**Figure 4** presents a visual representation of the differences between the two networks. The upper left histogram delineates the distribution of link lengths within the HPMS network across the United States, whereas the lower left histogram illustrates the distribution of link lengths within the FAF network across the same geographical scope. Note, in this study, most links of the



HPMS network were under 1 mile, whereas a significant proportion of links within the FAF network was more than 1 mile. The map displayed on the right of **Figure 4** exemplifies a typical instance of the divergence between the two networks. The green dots symbolize the end nodes of the TMAS network's links, whereas the blue dots signify those of the FAF network. Note, the HPMS network (green dots and links) was deliberately partitioned into smaller units. Consequently, a solitary FAF link may correspond to multiple HPMS links in practical application. In summary, the FAF network exhibited greater comprehensiveness with a diminished quantity of links but featured geometries that align more closely with real-world scenarios.

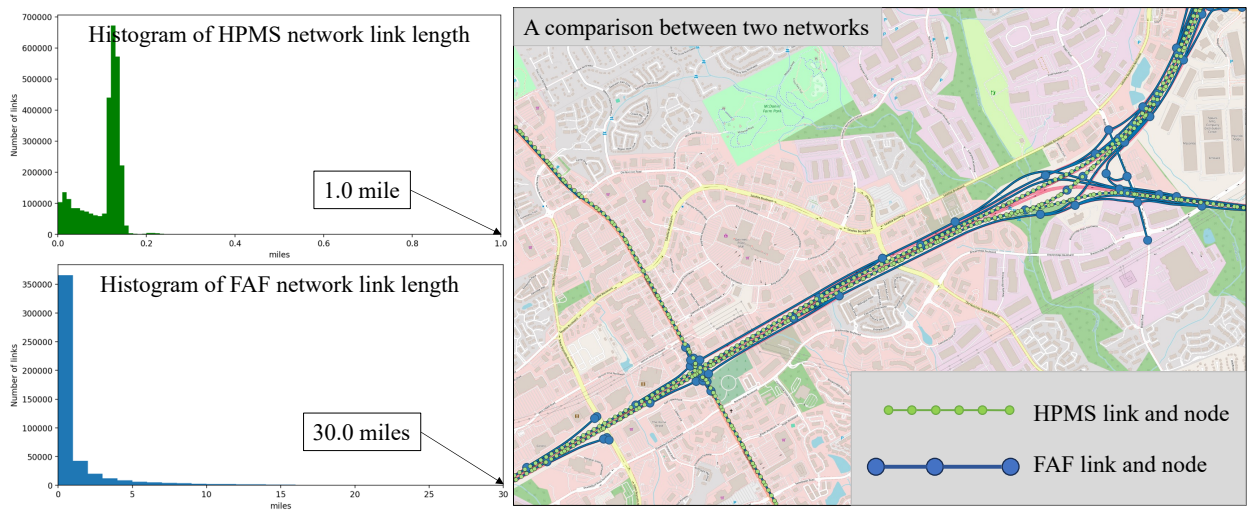

**Figure 4. A comparison of two networks: HPMS and FAF. FAF networks are more comprehensive and tailored for freight traffic analysis.**

**Table 1** further compares the disparities between the FAF network and the HPMS network regarding the quantities of links and nodes, the distribution of link lengths, and other parameters. Note, the median distance among FAF links measured 0.376 miles, whereas the median distance for HPMS links was markedly lower at 0.121 miles. Consistent with the observations, as shown in **Figure 4**, the FAF network was more comprehensive compared with its HPMS counterpart.



**Section 3.2** provides further discussion on the procedural intricacies involved in aligning the HPMS network with the FAF network.

**Table 1. The overall structure of the network**

| Network | Number of links | Total length (miles) | 25% link length | 50% link length | 75% link length |
|---|---|---|---|---|---|
| HPMS (National Highway System) | 3,017,746 | 377,673 | 0.078 | 0.121 | 0.132 |
| FAF | 487,394 | 766,646 | 0.161 | 0.376 | 0.997 |

The second phase of the analysis was dedicated to obtaining aggregated TMAS data. Given that the TMAS dataset encompasses records for every hour throughout the year, researchers can specify their time window of interest before conducting aggregation. **Table 2** describes several typical temporal intervals of interest. For instance, analysts may opt to scrutinize monthly traffic patterns by aggregating data monthly. Similarly, aggregation based on different times of the day or days of the week is feasible. Moreover, AADT values for trucks can be approximated by aggregating data spanning a full year. In essence, analysts can configure temporal intervals for aggregation in accordance with their research objectives.



**Table 2. Aggregation options involved in the study and provided by the analysis pipeline**

| Aggregation level | Description |
|---|---|
| Vehicle class | Aggregate probability by each vehicle class |
| All time | Aggregation observations over the whole year |
| Time of the day | Aggregate observations in different periods of the day; specifically, in this study:<br>1. T0: Observations between 22:00 and 4:00<br>2. T1: Observations between 4:00 and 10:00<br>3. T2: Observations between 10:00 and 16:00<br>4. T3: Observations between 16:00 and 22:00 |
| Weekday/weekend | Aggregate observations by weekdays and weekends |
| By month | Aggregate observations by different months |

Furthermore, the TMAS dataset encompasses traffic volume records categorized by vehicle class, enabling granular analysis at the level of individual vehicle classes. Within this study, the proportional representation of each vehicle class was carefully documented. Subsequently, employing spatial proximity and directional alignment, each TMAS station was precisely matched with the nearest link within the FAF network, a process elaborated on in **Section 3.2**.

In the third phase, the aggregated traffic volumes were converted into probabilities before the imputation process across the FAF network. Within the FAF network, the truck's AADT volume served as foundational data guiding the imputation and distribution of values throughout the network. **Section 3.3** explains the proposed algorithm in detail.

After running the imputation, Phase 4 involves a wide range of quantitative and qualitative tools to thoroughly analyze the advantages and limitations of the tool. Most of the experiments and evaluations were based on cross validation methods, as described in **Section 3.4**.



### 3.2 Estimating AADT values by matching HPMS links to FAF links

As outlined in **Section 3.1**, the imputation algorithm was exclusively executed within the framework of the FAF network because of its superior geometric properties and enhanced comprehensiveness. However, the AADT values derived from the HPMS dataset were contained within their own HPMS network. To transfer the AADT volume information from the HPMS network to the FAF network, the development of a matching algorithm was needed to align the respective outcomes.

Before the introduction of the matching algorithm, it was essential to assess the AADT value for each FAF link using AADT values derived from HPMS links. We observed the value against two networks (e.g., the right map of **Figure 4**), and **Figure 5** illustrates a representative case comparing two networks. As observed earlier, the FAF network delineates network geometries, whereas the HPMS network comprises numerous smaller segments with reduced geometric intricacies. Additionally, each HPMS link represents two directions unless the road is one way. Conversely, within the FAF network, each link of interstate highways solely represents traffic flow in a single direction.



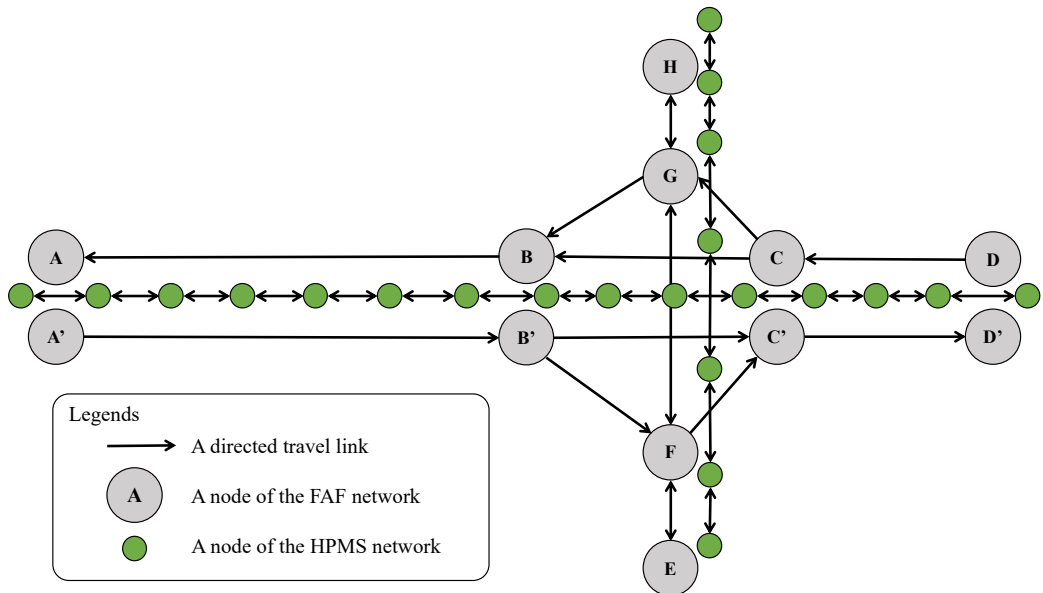

**Figure 5. A typical local representation of an interstate segment with an overpass arterial using FAF and TMAS networks.**

**Figure 6** shows the six main steps for aligning HPMS links with FAF links. First, both networks were converted into directed graphs, wherein each edge represented a single travel direction. Given that AADT quantifies traffic volume bidirectionally, the metric was halved to reflect unidirectional traffic volume. Subsequently, a composite series of geographical operations was performed to ensure the fidelity of AADT matching outcomes.

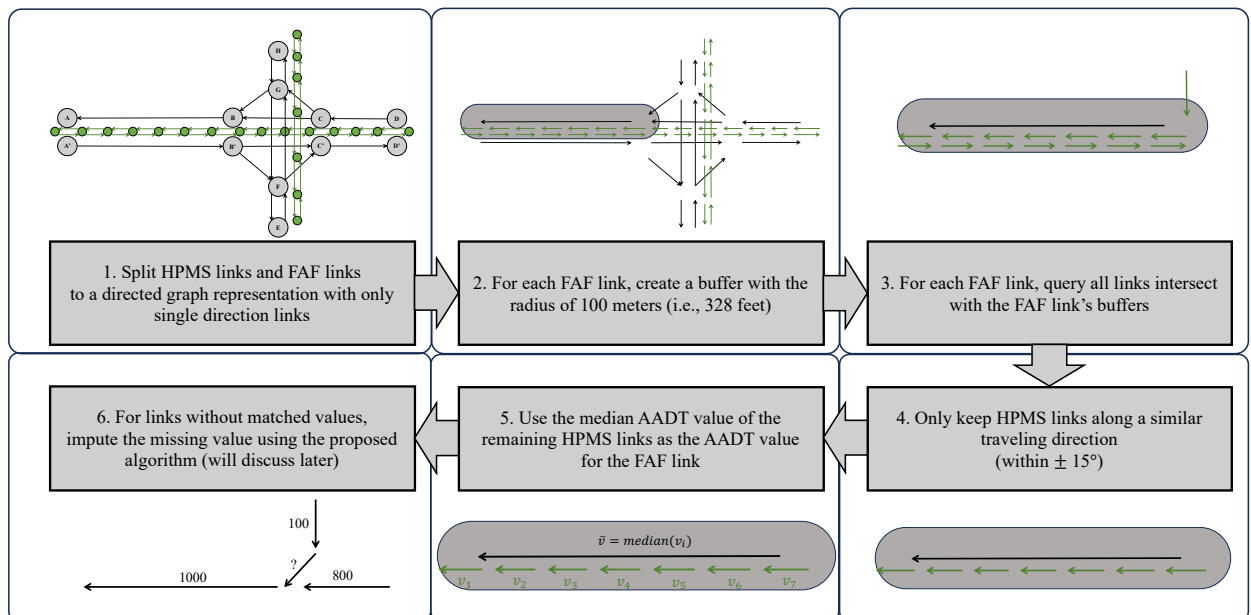

**Figure 6. The procedure of estimating AADT values from HPMS networks.**



The second stage allowed for creation of a buffer zone around the HPMS link, and all intersecting HPMS edges were systematically collated for each FAF link buffer. Subsequently, during Stage 4, only those HPMS edges demonstrating approximate alignment with the prevailing traveling direction (±15°) were retained. In Stage 5, the median HPMS value served as the reference HPMS value. Finally, in cases of FAF links without corresponding HPMS links, volume information was effectively imputed using the proposed algorithm.

Upon generating precise estimates of AADT for each FAF edge, data generated from TMAS stations were matched to the most reasonable FAF edge, considering spatial proximity and travel direction. **Figure 7** illustrates the process of aligning a TMAS station with FAF network links. It was imperative to discern between the two opposing travel directions, even though they shared identical TMAS station identifiers, and align them with their respective directional links. After the procedures of AADT estimation and the spatial alignment of TMAS stations with the nearest links, the FAF network comprehensively incorporated all pertinent information, encompassing AADT values, TMAS data recordings, and other relevant metrics. Because the algorithm was running at a nationwide scale and the density of detectors was low, an edge could be used to store information for TMAS stations at a fixed point.



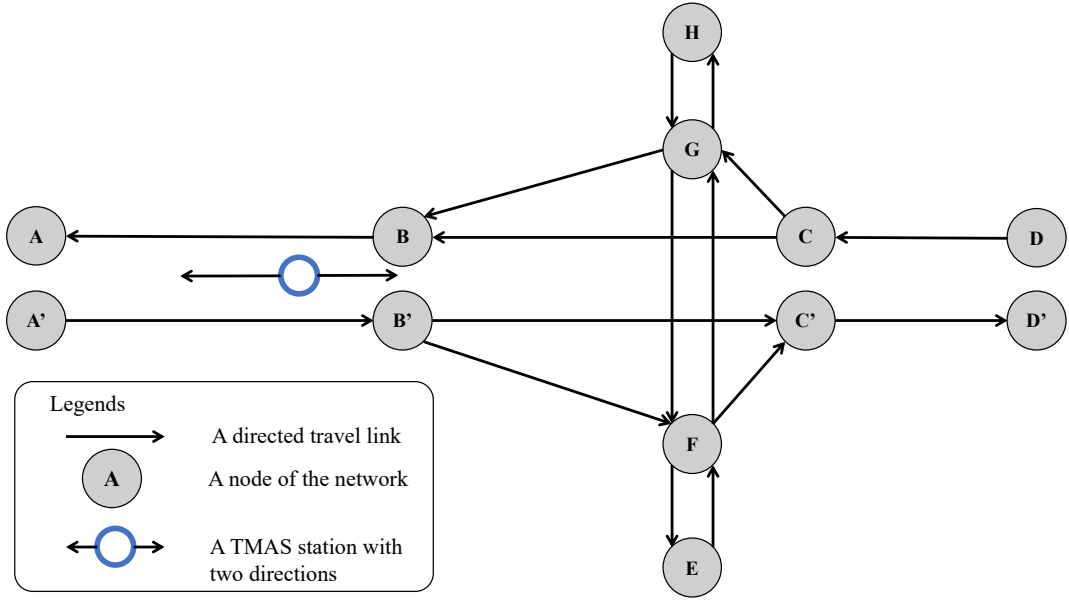

**Figure 7. A typical local representation of a highway segment with an overpass arterial network. A TMAS station (the blue circle) is located along the interstate highway.**

### 3.3 Imputing values over the network with prior weights

As described earlier, the FAF network was converted into a directional graph; each link corresponded to a specific travel direction, and each node established connections with two or more links to constitute a traffic network. Following the construction of such a network, values were interactively imputed across the network, as illustrated in **Figure 8**. The link was designated as the target for estimating imputed values. Initially, all incoming links to node $C$ (i.e., $\overrightarrow{AC}$ and $\overrightarrow{BC}$) and outgoing links from node $D$ (i.e., $\overrightarrow{DE}$ and $\overrightarrow{DF}$) were identified as neighboring links of $\overrightarrow{CD}$ (depicted in red in **Figure 8**). Subsequently, the values of $\overrightarrow{CD}$ were determined as the weighted average of the neighboring links, following the formula

$$\hat{y}_i = \frac{\sum_j w_j \hat{y}_j}{\sum_j w_j}, \forall\, y_j \in \{y_j : L_i \text{ and } L_j \text{ are neighboring links}\}, \tag{1}$$

where $L_i$ represents the target link for imputation (e.g., $\overrightarrow{CD}$); $L_j$ denotes the neighboring links of $L_i$; and $\hat{y}_i$ and $\hat{y}_j$ signify the information to be imputed or estimated within this context,



respectively. In practical application, $\hat{y}_i$ could signify the probability associated with a specific vehicle class, such as the probability of Vehicle Class 5, or represent traffic volume. Furthermore, $w_j$ refers to the weight attributed to each neighboring link. In the framework of this investigation, $w_j$ represents the truck AADT value of a link, derived through the methodology outlined in **Figure 6**.

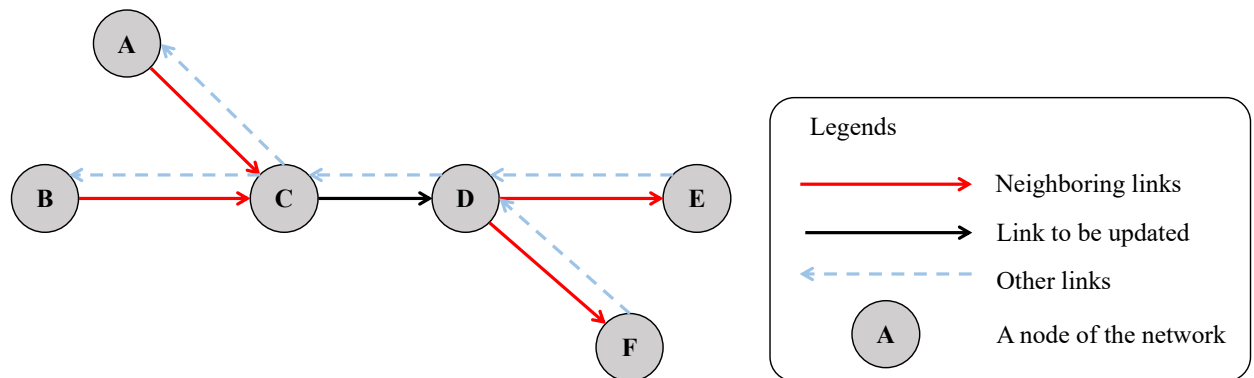

**Figure 8. Graph representing a local road segment to demonstrate the neighboring links (in solid red) of a link of interest (in solid black).**

Consequently, when imputing a probability distribution across the network, the target link's value is updated as a function of the weighted average of neighboring links, with weights determined by truck AADT volume. For the final step outlined in **Figure 6**, in instances where AADT values were either missing or deemed unrealistic, a similar imputation approach was employed, using the truck volume of the FAF network as link weights to estimate their AADT volumes, given existing neighboring links.

In addition to Equation (1), which serves as the foundational framework for imputing traffic values, the algorithm operated by systematically enumerating each link within the network, incrementally assigning values to links in a stepwise manner. Through repetition of this process, the algorithm converged toward a state in which link volumes approach equilibrium. Drawing



inspiration from the renowned shortest path searching methodology elucidated in the Bellman–Ford algorithm (Dasgupta and Papadimitriou, 2006), a novel approach for traffic imputation was introduced, accompanied by the corresponding pseudocode presented in **Figure 9**.

---
**Algorithm 1:** Iteratively imputing traffic information over the network
---
**Input:** Graph $G$, which contains AADT values for each edge $e$, observed link values for the corresponding edges; $K$ is the number of epochs to impute data
**Output:** imputed values for all links in $G$
1: For $i$ in $\{1, …, K\}$:
2: For each edge $e$ in $G$:
3: if $e.y$ is observed value:
4: continue
5:     $ns$ = get_neighboring_edges($e, G$)
6: if valid_to_impute($e, ns, G$):
7:     $e.y$ = impute_edge($e, ns, G$)
---

**Figure 9. The pseudocode of the graph algorithm running the imputation process.**

The algorithm abstained from imputing a value if it was already an observed value corresponding to a TMAS station, according to the condition "if $e.y$ is observed value." The parameter $K$ regulated the number of iterations the algorithm undergoes. The "valid_to_impute" function in Line 6 of the pseudocode described two exceptional scenarios in which imputation could not proceed until all links had been assigned values either upstream or downstream. One such scenario involved a merging link with multiple inbound links but only one outbound link, as depicted in **Figure 10**. The other scenario entailed multiple outbound links but only one inbound link. These instances represent specialized cases in which imputation is deferred until either all inbound (upstream) links or outbound (downstream) links are resolved. Note, all algorithms employed in this study are implemented using Python, whereas the graph of the FAF network leverages the NetworkX package (Hagberg et al., 2008).



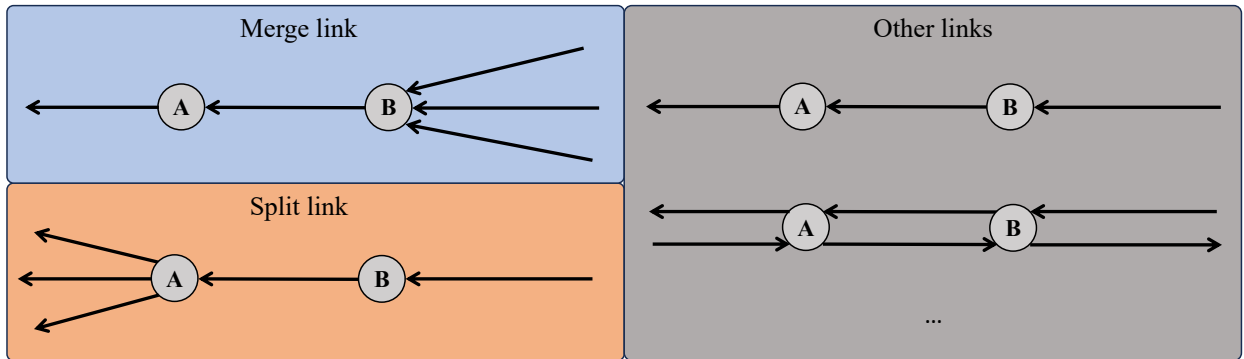

**Figure 10. Two special kinds of traffic links with special rules for imputation.**

### 3.4 Cross validation and evaluation metrics

Quantitative methodologies are essential for assessing the efficacy of the proposed algorithm. Specifically, a tenfold cross validation technique was employed for model evaluation. **Figure 11** describes the steps involved in the cross-validation process to assess the algorithm's performance. All TMAS stations were randomly partitioned into 10 subsets. Within each iteration of imputation, one data subset comprising 10% of TMAS stations was masked (depicted by blue rectangles in the figure). The remaining 90% of detectors were used to propagate values across the network. Performance was gauged by comparing the observed and predicted values for the masked validation set. To evaluate the model performance, four primary metrics were used: mean absolute error (MAE), root mean squared error (RMSE), Pearson's R2 values, and cross entropy loss (CEL). Additionally, further visual investigation tools, such as scatterplots and probability distribution maps, were employed to comprehend the model performance and the algorithm's behavior.



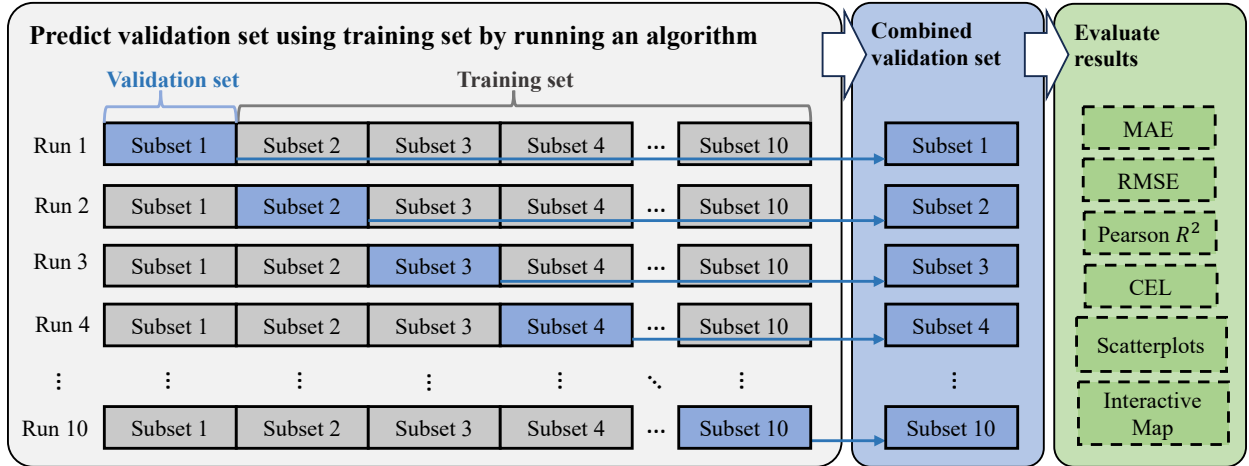

**Figure 11. A visualization of algorithm evaluation using cross validation.**

Three metrics were applied to evaluate the performance of traffic volume, including MAE, RMSE, and Pearson R2 values. These three metrics are defined as

$$MAE = \frac{\sum_{i=1}^{N}|y_i - \hat{y}_i|}{N}, \tag{2}$$

$$RMSE = \sqrt{\frac{\sum_{i=1}^{N}(y_i - \hat{y}_i)^2}{N}}, \text{ and} \tag{3}$$

$$R^2 = \rho^2_{y_i,\hat{y}_i} = \left(\frac{cov(y_i,\hat{y}_i)}{\sigma_{y_i}\sigma_{\hat{y}_i}}\right)^2, \tag{4}$$

where $N$ is the number of records and $y_i$ and $\hat{y}_i$ are the observed and predicted values, respectively. Meanwhile, $cov(y_i, \hat{y}_i)$ is the covariance between two series of data, and $\sigma_{y_i}$ and $\sigma_{\hat{y}_i}$ correspond to the standard deviation of observed (i.e., true) values and predicted values, respectively.

In addition to the three metrics in Equations (2)–(4), a fourth metric, CEL, was used to quantify the disparities between the predicted and actual observed distribution of vehicle classes. In this study, CEL was used to measure the distance between two distributions:



$$CEL(f, g) = -\sum_{i=5}^{13} f(i = c) \log(g(x = i)) \tag{5}$$

where $f(i)$ and $g(i)$ are the observed and expected probability distributions for different vehicle type $i$s, respectively.

## 4. Results

### 4.1 Data preparation: data selection and data quality check

To investigate the proposed method, our analyses relied on the 2021 TMAS dataset. For simplicity, only the class dataset was considered in this study, whereas the weight and volume datasets were excluded from our current framework. Consequently, all findings presented here are predicated solely on the class dataset.

A quality assessment was conducted to evaluate the availability and variability of the TMAS class dataset throughout the year, as depicted in **Figure 12**. The red dashed line delineates the average percentage of hours each state reported data per month. Most states exhibited a relatively high and consistent percentage (above 75%) of records generated, with the exception of South Dakota.



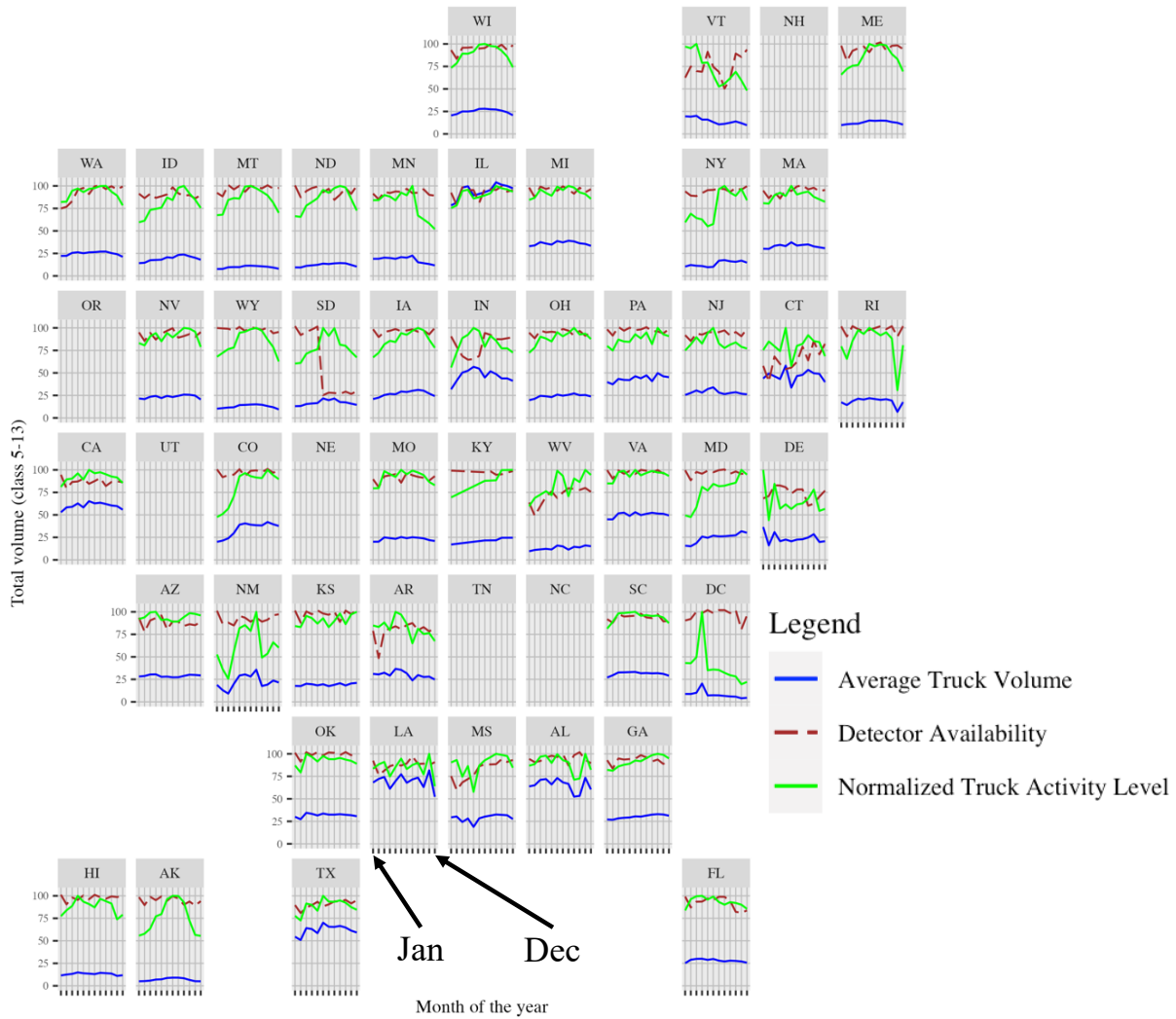

**Figure 12. Quality check of detector availability, average truck volume (normalized between 0 and 1), and truck activity level for each month for each state.**

Concurrently, the blue line in **Figure 12** illustrates the average truck volume normalized by the number of TMAS stations, with a scale ranging from 0 to 100 vehicles per hour. Illinois exhibited the highest observed traffic volume throughout the year, followed by Louisiana, Texas, Alabama, and California, indicating the operational intensity of each station on average. The volume was standardized from 0% to 100% to facilitate comparison of variations across different months, denoted by the green line. The busiest month was designated as 100%. Although many



northern states displayed a seasonal trend with significantly reduced traffic volume during winter months, the overall trend remained relatively stable throughout 2021.

TMAS data quality should be checked, and the volume inputs for the FAF network's edge weights should be reasonable. **Figure 13** shows the AADT truck volume over the maps for Texas and Georgia. The color bar represents the truck AADT volume ranges from 0 to 10,000. A visual check over different places showed that the truck AADT flow trend was reasonable.

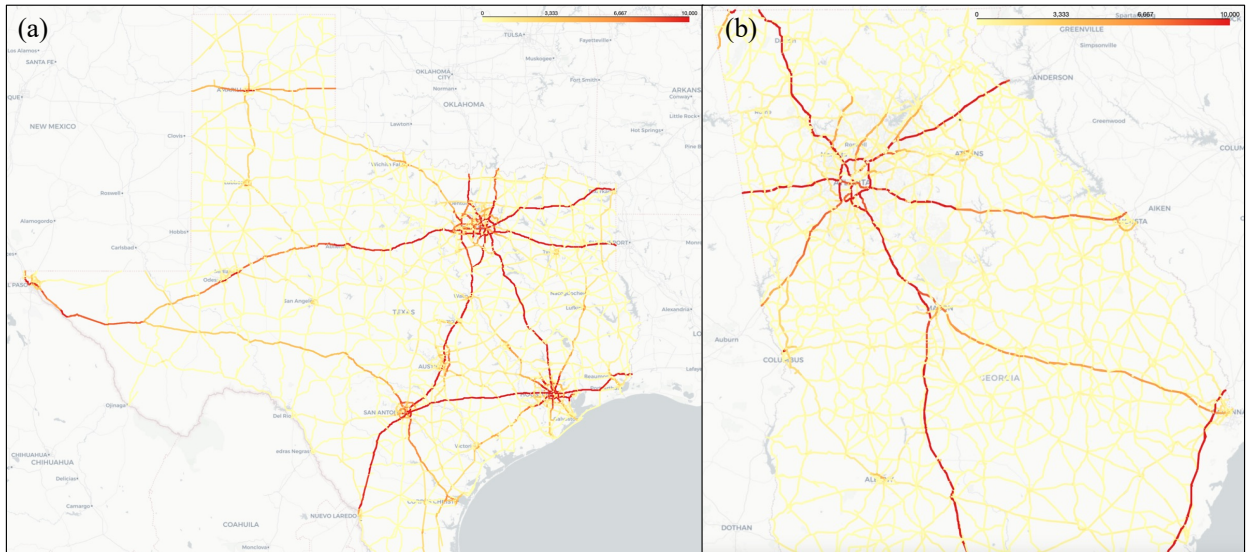

**Figure 13. A visualization of directional AADT of trucks for (a) Texas and (b) Georgia. The AADT values are matched to FAF links from HPMS links.**

### 4.2 Impute traffic volume by each traffic class for each TMAS station

As previously discussed, a tenfold cross validation procedure was conducted to evaluate the model. In each iteration, 10% of TMAS stations were masked as validation sets. Subsequently, the imputed values corresponding to these validation sets were compared with the observed values for model assessment.



**Figure 14** depicts the scatterplot analysis of the TMAS detectors. For each plot, the x-axis represents the observed values, whereas the y-axis denotes the predicted values. A reference line at a 45° angle illustrates the ideal scenario where predicted values align perfectly with observed values. The right subplot presents the aggregate volume data for all vehicles, and all subplots are visualized using a logarithmic scale. The logarithmic scale manifested an ellipsoidal pattern around the 45° line, suggesting the method's viability. The nine smaller subplots on the left depict scatterplots for individual vehicle classes ranging from 5 to 13. Notably, most large vehicles were either Class 5 or Class 9. Class 5 corresponded to two-axle trucks, such as single-unit trucks, whereas Class 9 comprised five-axle trucks, such as single-trailer trucks.

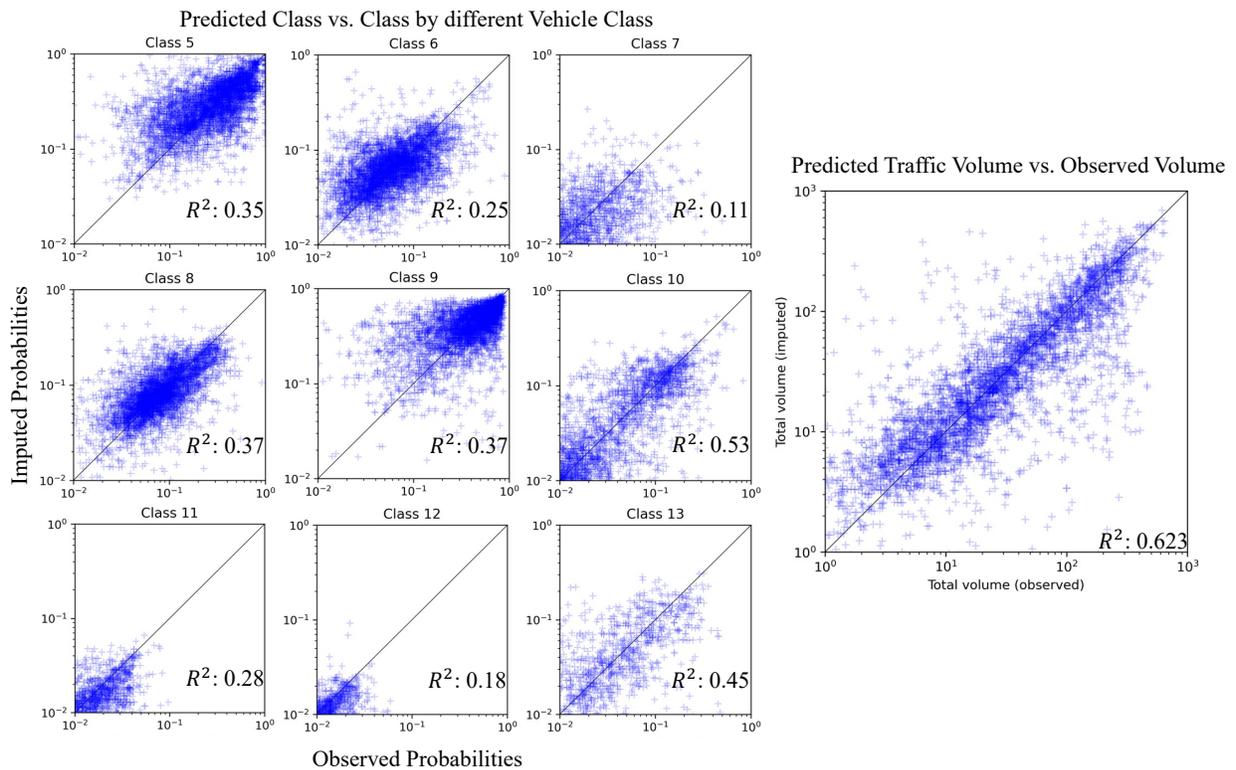

**Figure 14. Analysis of the $R^2$ of detectors of traffic volume from Vehicle Classes 5–13.**



**4.3 Breakdown traffic class distribution by different states**

To comprehensively assess outcomes across various vehicle classes, it is advantageous to disaggregate the findings based on geographic locations, given that each state contributes its distinct dataset to the TMAS database. **Figure 15** illustrates this disaggregation, presenting results segmented by individual states through a geo-facet plot. Each subplot corresponds to a specific state, identified by its abbreviation at the top. Within each subplot, three horizontal bars are depicted vertically in blue, green, and red, representing the R2 values for all vehicle classes, the MAE for Class 5, and the MAE loss for Class 9, respectively. States shown without data did not submit information to the TMAS dataset. Additionally, several states—including New Mexico, Kansas, Alabama, and the District of Columbia—demonstrated comparatively poorer performance than others.



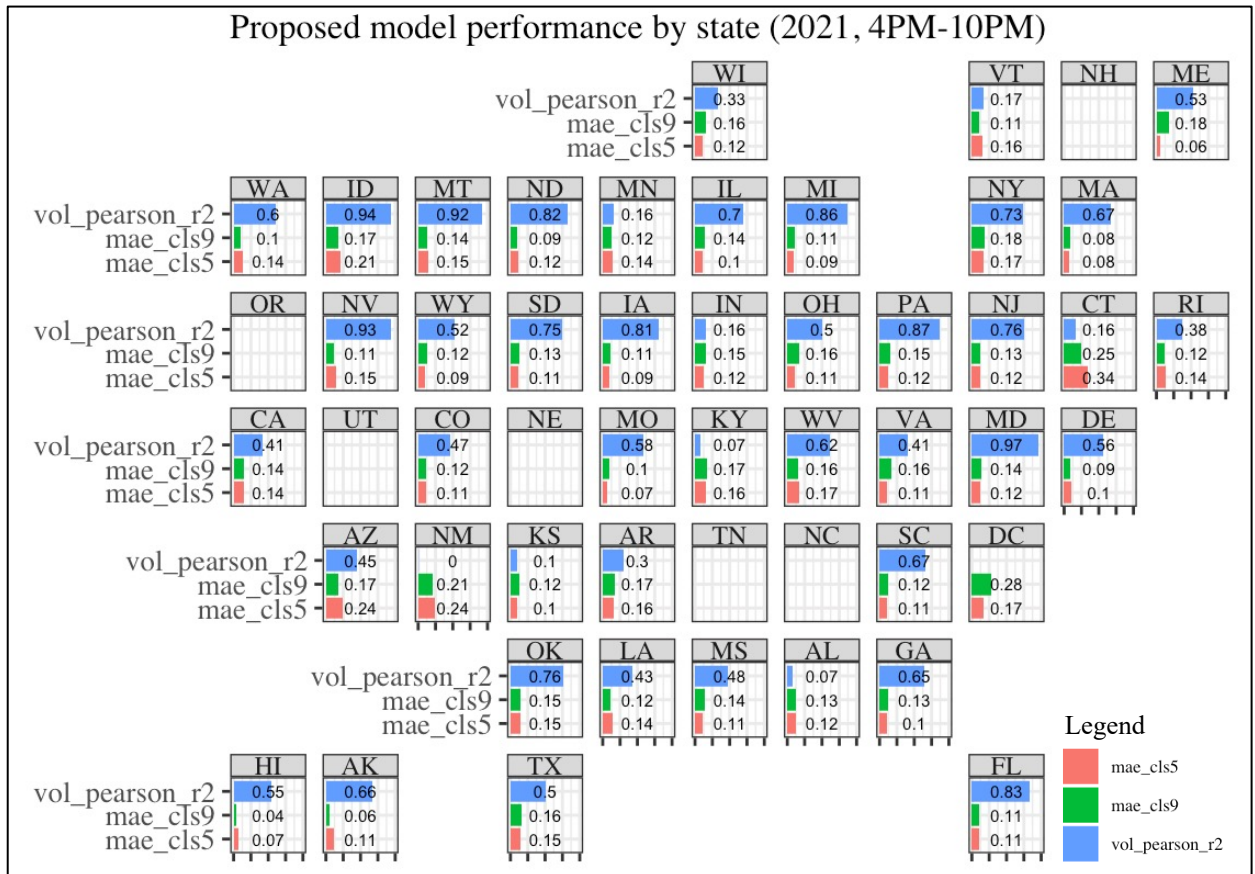

**Figure 15. Breakdown analysis for the performance of truck volume prediction for different states.**

In addition to the regional analysis, we investigated the performance variation across different times of the day.

**Table 3** presents a breakdown of performance metrics for truck volume classes across distinct time intervals. Each day was segmented into four periods: T1 (22:00–4:00), T2 (4:00–10:00), T3 (10:00–16:00), and T4 (16:00–22:00). Three metrics were reported: Pearson R2, MAE, and RMSE. Two other basic benchmark methods, GWR and Kriging, were compared with the proposed method. The results showed that the Kriging method has a poor performance with R2 ranges from 0.10 to 0.20. GWR and the proposed method had a similar output. For GWR, the performance was good because AADT was well correlated to the imputed traffic volume.



**Table 3. A comparison of different experiments segmented by time of the day (for weekdays)**

| Aggregation level | Average hourly traffic volume (Classes 5–13) | | | | | | | | |
|---|---|---|---|---|---|---|---|---|---|
| | GWR | | | Kriging | | | Proposed method | | |
| | $R^2$ | MAE | RMSE | $R^2$ | MAE | RMSE | $R^2$ | MAE | RMSE |
| All time | 0.59 | 28.48 | 65.59 | 0.16 | 58.73 | 96.30 | 0.57 | 15.67 | 42.26 |
| T1: (22:00–4:00) | 0.49 | 15.65 | 42.70 | 0.13 | 31.41 | 57.35 | 0.52 | 14.85 | 39.58 |
| T2: (4:00–10:00) | 0.61 | 31.27 | 68.80 | 0.20 | 61.81 | 99.85 | 0.58 | 32.05 | 70.93 |
| T3: (10:00–16:00) | 0.62 | 42.10 | 89.40 | 0.19 | 84.93 | 132.98 | 0.65 | 47.35 | 103.12 |
| T4: (16:00–22:00) | 0.59 | 29.53 | 65.61 | 0.19 | 35.59 | 58.36 | 0.52 | 30.22 | 71.78 |
| January 2021 | 0.66 | 23.62 | 51.60 | 0.16 | 53.53 | 86.50 | 0.57 | 26.18 | 57.78 |
| February 2021 | 0.48 | 27.60 | 79.23 | 0.15 | 57.15 | 93.78 | 0.52 | 28.84 | 66.17 |
| March 2021 | 0.54 | 27.95 | 73.75 | 0.17 | 59.04 | 98.90 | 0.55 | 28.57 | 67.43 |
| April 2021 | 0.56 | 28.71 | 68.89 | 0.16 | 59.95 | 97.58 | 0.54 | 29.63 | 67.60 |
| May 2021 | 0.47 | 29.26 | 78.41 | 0.17 | 56.73 | 94.34 | 0.52 | 28.69 | 67.62 |
| June 2021 | 0.48 | 30.66 | 81.17 | 0.18 | 60.64 | 97.61 | 0.51 | 31.38 | 70.96 |
| July 2021 | 0.59 | 28.65 | 63.55 | 0.16 | 60.92 | 96.28 | 0.53 | 29.36 | 67.18 |
| August 2021 | 0.57 | 29.03 | 68.83 | 0.16 | 61.34 | 99.51 | 0.52 | 30.51 | 69.96 |
| September 2021 | 0.71 | 27.00 | 52.46 | 0.18 | 59.69 | 93.06 | 0.58 | 30.03 | 62.33 |
| October 2021 | 0.70 | 26.57 | 53.66 | 0.15 | 60.92 | 96.12 | 0.61 | 28.58 | 60.40 |
| November 2021 | 0.71 | 26.25 | 52.20 | 0.18 | 59.27 | 93.98 | 0.63 | 28.67 | 59.12 |
| December 2021 | 0.67 | 26.03 | 54.42 | 0.20 | 56.48 | 89.92 | 0.63 | 27.31 | 57.04 |

Despite initial concerns that temporal segmentation might adversely affect performance, the Pearson R2 outcomes exhibited only marginal degradation for the proposed method. Instead of using a statistical regression method, the proposed method gets a similar result solely based on a physical informed algorithm.

**Table 4** disaggregates the performance by each state. Five temporal aggregation levels were reported: all time, T1, T2, T3, and T4. The RMSEs for Classes 5 and 9 were examined. Among the distribution of Classes 5–13, the CEL was reported to measure the discrepancies between observed and predicted class distributions.



**Table 4. A summary of cross validation experiment results of the vehicle class shares of each state**

| State name | Vehicle class probability (proposed method) | | | | | | | | | | | | | | |
|---|---|---|---|---|---|---|---|---|---|---|---|---|---|---|---|
| | Class 5 (RMSE) | | | | | Class 9 (RMSE) | | | | | All classes (CEL) | | | | |
| | All | T1 | T2 | T3 | T4 | All | T1 | T2 | T3 | T4 | All | T1 | T2 | T3 | T4 |
| All states | .039 | .042 | .037 | .036 | .047 | .040 | .042 | .038 | .039 | .048 | 1.442 | 1.446 | 1.471 | 1.444 | 1.370 |
| Alabama | .119 | .026 | .020 | .017 | .024 | .137 | .032 | .020 | .020 | .026 | 1.517 | 1.44 | 1.470 | 1.469 | 1.430 |
| Alaska | .165 | .066 | .061 | .039 | .042 | .084 | .034 | .020 | .010 | .009 | 1.320 | 1.575 | 1.398 | 1.311 | 1.212 |
| Arizona | .255 | .186 | .117 | .108 | .106 | .185 | .117 | .063 | .058 | .068 | 1.323 | 1.653 | 1.395 | 1.370 | 1.221 |
| Arkansas | .155 | .026 | .030 | .026 | .041 | .174 | .043 | .029 | .038 | .055 | 1.364 | 1.433 | 1.358 | 1.390 | 1.355 |
| California | .152 | .044 | .036 | .036 | .054 | .148 | .039 | .037 | .038 | .049 | 1.201 | 1.306 | 1.271 | 1.205 | 1.137 |
| Colorado | .123 | .022 | .027 | .020 | .030 | .137 | .029 | .029 | .029 | .032 | 1.560 | 1.475 | 1.572 | 1.524 | 1.525 |
| Connecticut | .275 | .066 | .105 | .095 | .135 | .196 | .096 | .070 | .069 | .066 | 1.816 | 1.924 | 1.848 | 1.799 | 1.792 |
| Delaware | .180 | .034 | .028 | .058 | .054 | .092 | .032 | .017 | .013 | .022 | 1.474 | 1.879 | 1.471 | 1.486 | 1.337 |
| District | .215 | .003 | .059 | .091 | .002 | .054 | .018 | .006 | .003 | .002 | 1.694 | 1.963 | 2.979 | 1.844 | 0.872 |
| Florida | .121 | .028 | .022 | .024 | .034 | .137 | .03 | .031 | .032 | .040 | 1.377 | 1.342 | 1.44 | 1.395 | 1.266 |
| Georgia | .104 | .020 | .024 | .025 | .032 | .143 | .028 | .039 | .041 | .045 | 1.366 | 1.32 | 1.367 | 1.343 | 1.335 |
| Hawaii | .066 | .011 | .008 | .008 | .006 | .040 | .005 | .003 | .002 | .001 | 0.834 | 0.877 | 1.077 | 0.869 | 0.505 |
| Idaho | .233 | .072 | .074 | .058 | .088 | .218 | .058 | .046 | .049 | .070 | 1.843 | 1.958 | 1.878 | 1.823 | 1.797 |
| Illinois | .118 | .017 | .020 | .020 | .033 | .159 | .032 | .033 | .036 | .050 | 1.308 | 1.251 | 1.316 | 1.302 | 1.294 |
| Indiana | .126 | .025 | .026 | .024 | .045 | .154 | .043 | .039 | .045 | .068 | 1.286 | 1.217 | 1.327 | 1.304 | 1.253 |
| Iowa | .101 | .008 | .018 | .013 | .026 | .118 | .008 | .023 | .022 | .031 | 1.239 | 1.058 | 1.274 | 1.241 | 1.196 |
| Kansas | .104 | .009 | .018 | .017 | .026 | .110 | .010 | .021 | .019 | .024 | 1.213 | 1.197 | 1.259 | 1.219 | 1.198 |
| Kentucky | .207 | .111 | .055 | .055 | .098 | .215 | .076 | .070 | .063 | .097 | 1.469 | 1.56 | 1.504 | 1.502 | 1.406 |
| Louisiana | .150 | .065 | .041 | .047 | .042 | .109 | .067 | .045 | .062 | .039 | 1.259 | 2.126 | 1.264 | 1.331 | 1.238 |
| Maine | .050 | .003 | .006 | .007 | .007 | .159 | .060 | .028 | .034 | .043 | 1.613 | 1.45 | 1.634 | 1.659 | 1.544 |
| Maryland | .120 | .041 | .024 | .027 | .054 | .153 | .042 | .032 | .039 | .055 | 1.424 | 1.365 | 1.48 | 1.485 | 1.302 |
| Massachusetts | .078 | .010 | .012 | .018 | .019 | .085 | .013 | .014 | .018 | .023 | 1.540 | 1.483 | 1.527 | 1.522 | 1.531 |
| Michigan | .097 | .012 | .013 | .010 | .023 | .133 | .020 | .022 | .024 | .036 | 1.478 | 1.404 | 1.543 | 1.452 | 1.387 |
| Minnesota | .146 | .034 | .032 | .035 | .050 | .143 | .043 | .024 | .036 | .038 | 1.589 | 1.621 | 1.529 | 1.565 | 1.557 |
| Mississippi | .140 | .031 | .036 | .032 | .049 | .178 | .052 | .062 | .058 | .066 | 1.810 | 1.599 | 1.595 | 1.555 | 1.581 |
| Missouri | .070 | .017 | .015 | .008 | .021 | .094 | .022 | .021 | .015 | .025 | 1.215 | 1.203 | 1.262 | 1.237 | 1.166 |
| Montana | .161 | .039 | .038 | .036 | .058 | .150 | .030 | .033 | .030 | .042 | 1.658 | 1.654 | 1.711 | 1.666 | 1.645 |
| Nebraska | Data not available | | | | | | | | | | | | | | |



| State name | Vehicle class probability (proposed method) | | | | | | | | | | | | | | |
|---|---|---|---|---|---|---|---|---|---|---|---|---|---|---|---|
| | Class 5 (RMSE) | | | | | Class 9 (RMSE) | | | | | All classes (CEL) | | | | |
| | All | T1 | T2 | T3 | T4 | All | T1 | T2 | T3 | T4 | All | T1 | T2 | T3 | T4 |
| Nevada | .131 | .004 | .078 | .046 | .037 | .128 | .018 | .042 | .030 | .026 | 1.422 | 1.263 | 1.655 | 1.450 | 1.306 |
| New Hampshire | Data not available | | | | | | | | | | | | | | |
| New Jersey | .151 | .030 | .033 | .036 | .053 | .153 | .038 | .036 | .041 | .048 | 1.376 | 1.413 | 1.413 | 1.377 | 1.228 |
| New Mexico | .321 | .145 | .155 | .136 | .166 | .265 | .121 | .121 | .106 | .142 | 1.856 | 1.989 | 1.872 | 1.771 | 1.814 |
| New York | .156 | .046 | .045 | .047 | .066 | .184 | .054 | .051 | .065 | .080 | 1.640 | 1.668 | 1.701 | 1.686 | 1.599 |
| North Carolina | Data not available | | | | | | | | | | | | | | |
| North Dakota | .120 | .026 | .023 | .019 | .030 | .110 | .031 | .019 | .021 | .028 | 1.646 | 1.759 | 1.635 | 1.663 | 1.631 |
| Ohio | .120 | .018 | .028 | .025 | .042 | .160 | .039 | .045 | .046 | .065 | 1.425 | 1.396 | 1.461 | 1.402 | 1.402 |
| Oklahoma | .157 | .056 | .035 | .031 | .049 | .154 | .054 | .039 | .036 | .047 | 1.304 | 1.352 | 1.327 | 1.351 | 1.234 |
| Oregon | Data not available | | | | | | | | | | | | | | |
| Pennsylvania | .148 | .040 | .032 | .036 | .062 | .184 | .048 | .049 | .056 | .082 | 1.465 | 1.467 | 1.485 | 1.491 | 1.374 |
| Rhode Island | .169 | .039 | .025 | .036 | .037 | .148 | .018 | .024 | .032 | .023 | 1.246 | 1.412 | 1.266 | 1.251 | 1.181 |
| South Carolina | .130 | .034 | .029 | .030 | .035 | .130 | .031 | .032 | .032 | .033 | 1.259 | 1.238 | 1.276 | 1.264 | 1.157 |
| South Dakota | .108 | .008 | .021 | .019 | .022 | .143 | .010 | .029 | .023 | .026 | 1.472 | 1.396 | 1.531 | 1.496 | 1.421 |
| Tennessee | Data not available | | | | | | | | | | | | | | |
| Texas | .146 | .044 | .042 | .040 | .049 | .158 | .049 | .046 | .042 | .055 | 1.325 | 1.174 | 1.365 | 1.330 | 1.327 |
| Utah | Data not available | | | | | | | | | | | | | | |
| Vermont | .300 | .137 | .143 | .114 | .137 | .238 | .083 | .056 | .056 | .082 | 1.763 | 1.783 | 1.761 | 1.672 | 1.633 |
| Virginia | .118 | .021 | .024 | .029 | .036 | .170 | .035 | .042 | .048 | .051 | 1.368 | 1.279 | 1.382 | 1.387 | 1.293 |
| Washington | .148 | .056 | .049 | .049 | .060 | .110 | .031 | .018 | .023 | .029 | 1.758 | 1.823 | 1.827 | 1.789 | 1.609 |
| West Virginia | .184 | .084 | .049 | .070 | .079 | .190 | .077 | .057 | .070 | .066 | 1.437 | 1.416 | 1.52 | 1.504 | 1.315 |
| Wisconsin | .122 | .025 | .023 | .021 | .042 | .181 | .046 | .050 | .046 | .057 | 1.490 | 1.383 | 1.513 | 1.474 | 1.404 |
| Wyoming | .115 | .023 | .021 | .023 | .029 | .144 | .028 | .030 | .031 | .041 | 1.613 | 1.681 | 1.669 | 1.635 | 1.586 |



### 4.4 Model evaluation by visualizing imputation results

**Figure 16** illustrates the performance breakdown of traffic volume prediction across individual states, employing cross validation methodology for analysis. The x-axis denotes the density of TMAS stations, whereas the y-axis represents the Pearson R2 coefficient. Each data point corresponds to a state's prediction performance, with the size of the marker indicating the number of TMAS stations within the state. Some states are excluded from the graph because they do not have TMAS stations. Additionally, the District of Columbia and Hawaii were excluded from the analysis because they exhibited outlier characteristics owing to unique geometries or a scarcity of detectors. States with higher density seem to have more stable performances.

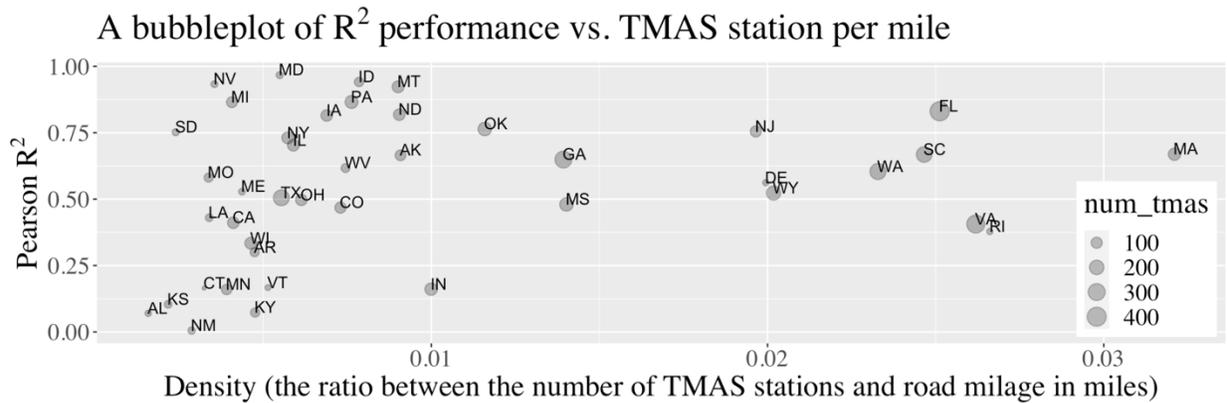

**Figure 16. A bubble plot comparing the Pearson $R^2$ performance for traffic volume with the density of TMAS stations by state.**

**Figure 17** illustrates a comparative analysis between imputed truck volume and HPMS truck volume. The left subplot displays the forecasted values generated by the proposed algorithm, whereas the right subplot exhibits the HPMS's truck AADT volume. Each blue point on the left subplot corresponds to a TMAS station. As anticipated, the imputation outcomes were similar to those of the historical AADT plot, despite AADT values being solely used as weights for



network links within the algorithm. This underscores the efficacy of the algorithm in accurately

imputing traffic volume across the network using the TMAS dataset.

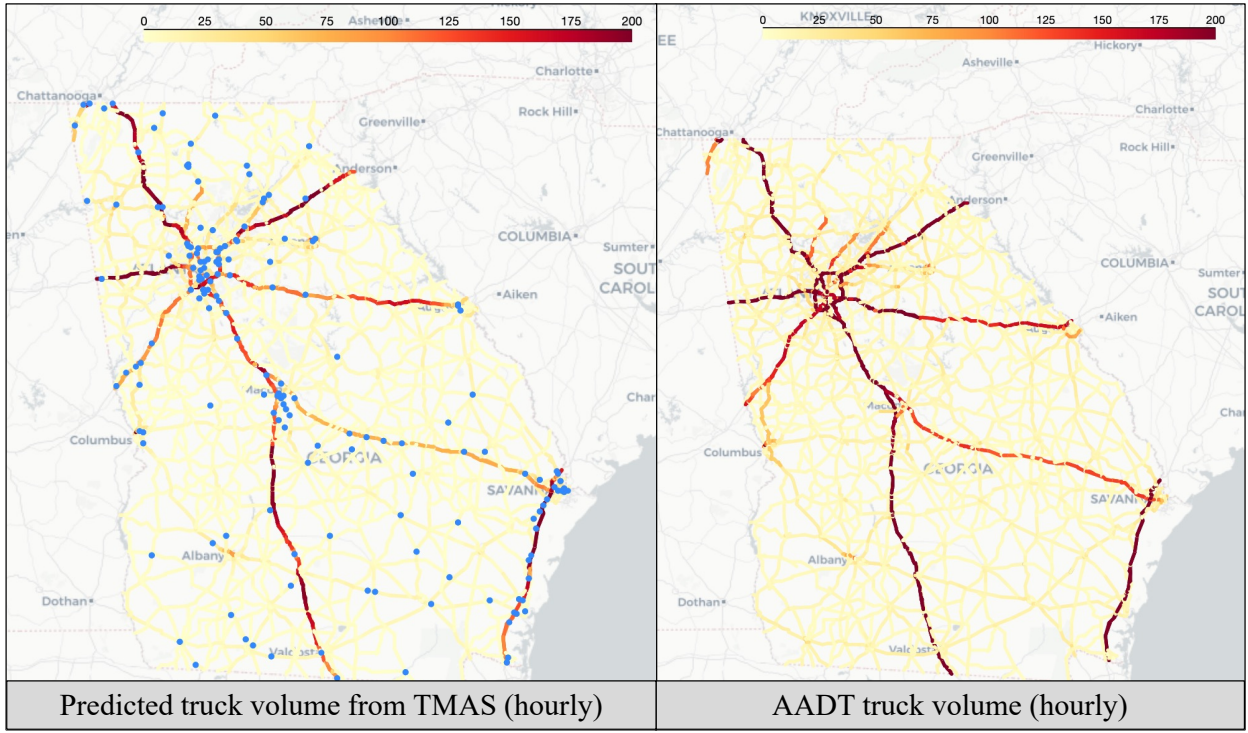

| Predicted truck volume from TMAS (hourly) | AADT truck volume (hourly) |

**Figure 17. A comparison between imputed truck value per hour and HPMS AADT truck volume (normalized to hourly volume) for Georgia.**

### 4.5 Analysis of truck travel pattern along highway: a preliminary case study of Savannah, Georgia

A significant advantage of the proposed method lies in its capacity to extrapolate values across

the network, transcending the confines of fixed TMAS stations. Furthermore, in addition to its

capability to analyze traffic patterns at national or state levels, the proposed method proves

highly valuable for scrutinizing local activities through the strategic placement of detectors along

traffic links. This spatial analysis is automatically facilitated by the imputation algorithm, which

operates seamlessly within the traffic network. To illustrate the efficacy of the tool, a case study

was conducted, focusing on Savannah, Georgia. By 2023, the port of Savannah had become the



fourth largest port in the United States (Lauriat, 2024), underscoring the importance of assessing truck performance along the major highways in proximity to Savannah.

**Figure 18** illustrates the distribution of vehicle classes across varying times of the day and on weekdays versus weekends. Analysis revealed that within the subset of Vehicle Classes 9–13, Class 9 vehicles (i.e., trailer trucks) are the most common vehicles traveling on the eastbound and westbound Interstate 16 highways. In contrast, the use of Interstate 95 by Class 9 vehicles was comparatively lower, with Class 5 (two-axle vehicles, such as delivery trucks) being more prevalent. The figure also depicts the volume of Class 9 vehicles. Integration of this data with speed profiles and emission models facilitated the estimation of emission effects on major roadway segments.



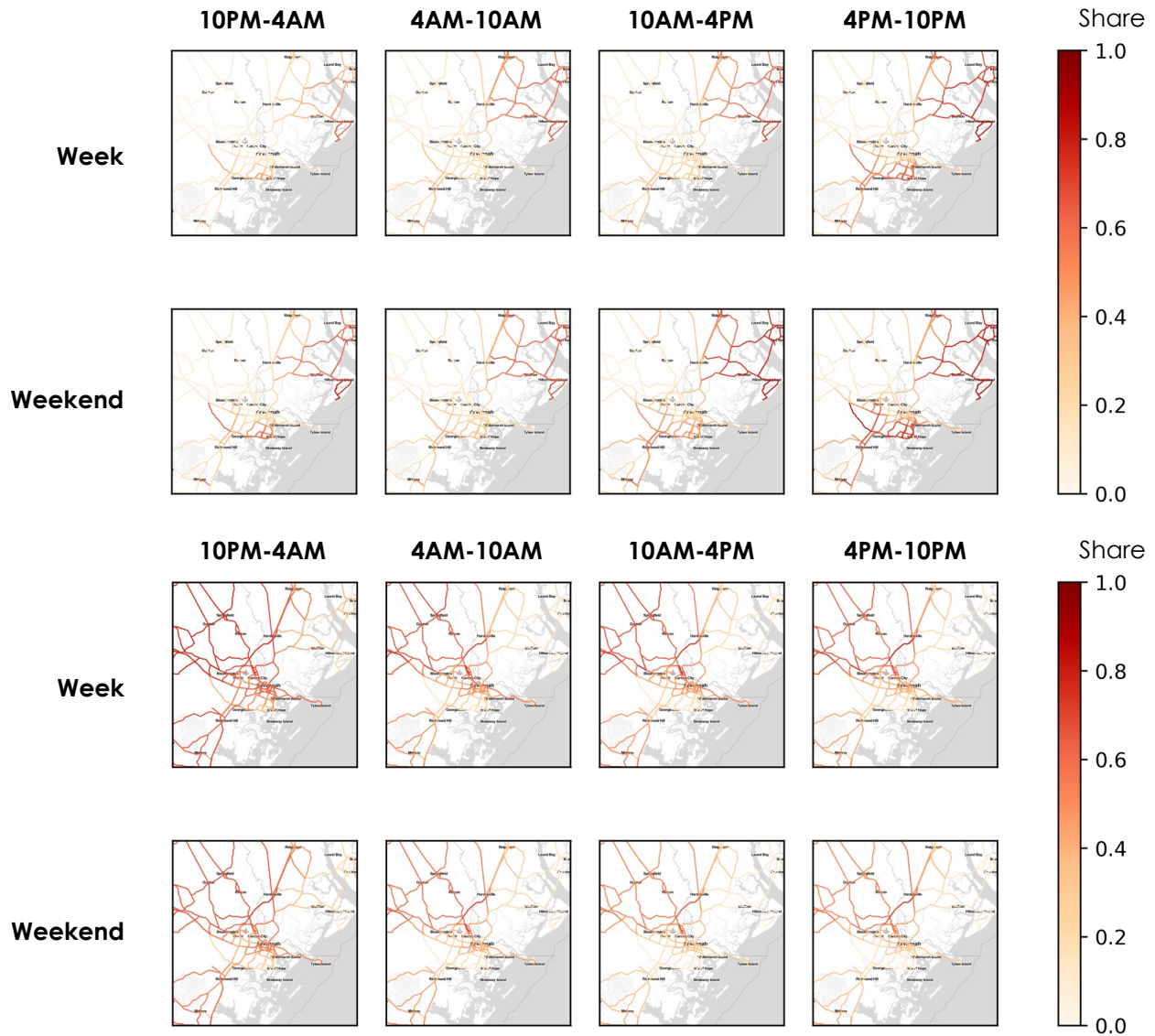

**Figure 18.** A visualization of the ratio of Classes 5 and 9 for different times of the day for weekdays and weekends in Savannah, Georgia.

### 4.6 A further visualization of specific stations: analyzing outliers

Assessing performance at the level of individual detectors for comprehensive analysis was imperative. **Figure 19** illustrates an interactive map using the Folium Python package, wherein each circle denotes a TMAS station. The size of each circle corresponds to the hourly truck traffic volume at the station, and the angle of each pie chart corresponds to a TMAS station. The angular sectors of the green and red pies represent the shares of Classes 5 and 9, respectively.



Smaller angles signify smaller share. Note that a TMAS station could have one or two directions. The figure therefore provides a comprehensive visualization of truck volume and class distributions. The figure also reveals that many TMAS stations were located at connector roads instead of along busy highways.

The map's interactivity enables users to click on circles to access the vehicle class distribution for each station. For example, the bottom row of **Figure 19** shows the class distributions of three corresponding TMAS stations. The blue line signifies the observed percentage for each class, whereas the orange line depicts the percentage predicted or imputed by the proposed algorithm. Researchers can use the chart to check the distribution of trucks over places to understand truck traffic patterns over the highway system.

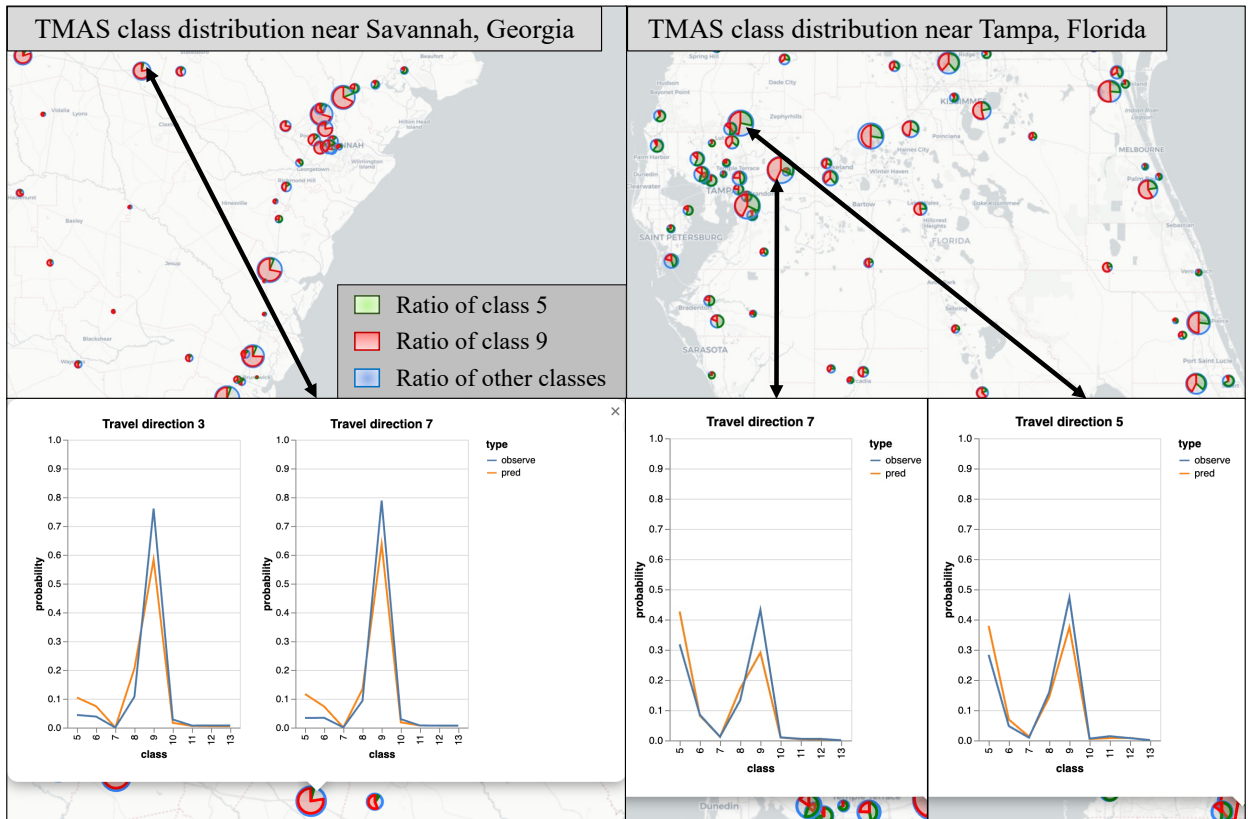

**Figure 19. An interactive map of TMAS stations visualizing truck volume and vehicle class distributions near (left) Savannah, Georgia, and (right) near Tampa, Florida.**



## 5. Conclusion

In this study, we introduced a novel iterative method for imputing traffic information across a traffic network. Beginning with a foundational HPMS dataset of AADT values, our approach enabled efficient imputation of observations at fixed locations throughout the traffic network. Unlike conventional interpolation techniques that primarily estimate AADT values, our method extends its capabilities to estimate aggregated traffic information across both temporal and spatial dimensions, accounting for diverse vehicle classes. The method's versatility and underlying analytical framework render it highly applicable in various domains, including traffic analysis, emission modeling, and truck behavior analysis. Notably, our study offers several new perspectives on a national scale, leveraging insights derived from the TMAS dataset, thus advancing the state-of-the-art traffic data analysis based on the TMAS dataset.

As a novel approach, the proposed method may have some inherent limitations. Despite enhancements to the quality of the TMAS dataset over the preceding 5 years, significant room for refinement persists. First, the integrity of certain state-contributed data remains suspect, with concerns arising with the potential use of synthesized data for estimating absent observations. Second, the heterogeneity of detector types and data formats across different states presents an obstacle to the conversion of records into the standardized TMAS format. Within the TMAS data schema, ambiguous descriptions remain. Notably, in the context of detector alignment, the dataset exclusively accommodates eight cardinal travel directions (i.e., east, west, north, south, northeast, northwest, southeast, and southwest), complicating station-to-direction alignment procedures. It is plausible, for instance, for a directional designation such as northbound to correspond to eastbound in practice, owing to overarching highway nomenclature, such as I-40 E. Furthermore, certain states have not submitted vehicle classification data, although it is likely



that many of these states possess the relevant data. The complexity inherent in the TMAS dataset's 13-class "F-scheme" may prove prohibitive, necessitating advocacy for states to furnish a simplified classification system when F-scheme is not possible. Such a simplification could entail the adoption of a truncated F-scheme featuring two to three most dominant vehicle classes, such as Class 5 for delivery trucks, Class 9 for single-trailer trucks, and Class 13 for double-trailer trucks.

Despite data quality considerations, the proposed methodology has several inherent limitations. Each computational step potentially can introduce errors. The estimation of AADT values for individual segments within the FAF network based on the HPMS network may incur inaccuracies because of the intricate geometries of the network. Additionally, in the process of imputation, the determination of the termination criterion necessitates manual intervention to regulate the parameter $K$, the number of epochs through the graph's edges. Furthermore, the imputation technique relies on the assumption of AADT truck volume, which commonly deviates from reality if investigated for some special time range.

Overall, our study demonstrates the feasibility of estimating freight traffic through the integration of AADT and TMAS observations, employing imputation techniques across the network. The sparse distribution of TMAS stations presents a significant challenge that has not been thoroughly investigated. This concept holds promise for further development and adaptation into alternative methodologies. For instance, there is potential to leverage deep learning techniques such as graph neural networks to facilitate information propagation across the network based on TMAS. Additionally, merging the three TMAS datasets—volume, class,



and weight—offers the prospect of establishing a cohesive digital twin for analyzing freight traffic dynamics.

## 6. Declaration of Generative AI and AI-Assisted Technologies in the Writing Process

During the preparation of this work, the authors used GPT-3.5 Turbo to improve the fluency and clarity of the writing. After using this tool, the authors reviewed and edited the content as needed and take full responsibility for the content of the publication.

## 7. Acknowledgements

The authors would like to thank the Federal Highway Administration, particularly Steven Jessberger, who approved the use of the internal TMAS dataset for this study for the student authors involved. The authors also gratefully acknowledge the funding and research opportunities provided by the US Department of Energy's Office of Energy Efficiency and Renewable Energy and the Graduate Advancement Training and Education program, facilitating collaborative research between the University of Tennessee, Knoxville, and the US Department of Energy's Oak Ridge National Laboratory.